\documentstyle[amsmath,amssymb,12pt,epsf]{article}
\def\half{\mbox{\small $\frac{1}{2}$}}

\def\ka{\kappa}

\begin{document}

\begin{flushright} 
Bicocca-FT-03-24\\
CU-TP-1095
\end{flushright}

\vskip50pt
\begin{center}
\begin{title}
\title{\large\bf BFKL Dynamics in Jet Evolution}\\

\vskip 10pt {G. Marchesini$^{\,a}$ and A.H.
  Mueller$^{\,b,}$\footnote{This research is supported
    in part by the US Department of Energy}\\
  \vskip 5pt {\small {\it$^a$ Dipartimento di Fisica,
      Universit\`a di Milano-Bicocca and \\
      INFN, Sezione di Milano, Italy\\
      $^b$ Department of Physics
      Columbia University\\
      New York, New York 10027, USA}}}
\end{title}
\end{center}
\vskip 10pt

\begin{flushleft}
 {\bf Abstract}
\end{flushleft}

\indent We calculate $e^+e^-\rightarrow Q\bar{Q}(k)+$ anything in a
certain momentum, $k,$ region for heavy quark-antiquark $(Q\bar{Q})$
production.  In our chosen region we find that the number of heavy
quark pairs produced is determined by BFKL dynamics and the energy
dependence of the number of pairs is given by $\alpha_P\!-\!1$ the hard
pomeron intercept.

\vskip 10pt
\noindent {\bf 1.  Introduction}
\vskip 6pt

\indent The topic of this paper, BFKL dynamics \cite{aev,sky} in QCD
jet evolution \cite{tto,DKMT} is perhaps surprising since it is
generally thought that BFKL dynamics has uniquely to do with
high-energy scattering and not at all to do with jet physics.  Indeed,
to a large extent jet evolution is an understood and well-tested
subject with the dominant dynamics being given by double logarithmic
perturbative terms coming from emission off the primary energetic
partons.  Their resummation leads to a Sudakov type of distribution.
Until very recently there has been no hint of BFKL dynamics in jet
physics.  However, about two years ago Dasgupta and Salam \cite{Das}
observed that in order to calculate certain ``non-global'' observables
such as the probability, $\Sigma,$ that a jet of energy $ E$ decay
while emitting less than a certain amount of energy, $E_{\rm out},$
into a specified angular region away from the jet, it is necessary to
evaluate single logarithmic terms in the jet evolution.  For $\Sigma$
double logarithmic terms, soft and collinear, cancel leaving only the
logarithms from soft non-collinear emissions originating from
successive branching of soft gluons at comparable angles.

Stimulated by the results of Dasgupta and Salam, Banfi, Marchesini and
Smye (BMS) \cite{Ban} found an equation for all the single logarithmic
terms contributing to $\Sigma$ derived from the multi-soft gluon
distributions given in Ref.~\cite{tto}.  The surprise is that the
dominant part of the BMS equation is identical to the Kovchegov
equation \cite{gov}.  This is surprising for several reasons: (i) The
Kovchegov equation is an approximate equation for dealing with small-$x$
evolution, or high-energy scattering, when unitarity limits are being
reached. The BMS equation is not an approximate equation; it is an
equation which exactly deals with single logarithmic terms. However,
it appears to have nothing to do with unitarity limits, rather, the
nonlinear aspects of the equation are related to the nature of the
non-global observable $\Sigma.$ It is due to the non linearity of
successive branchings of soft gluons (or jets) at comparable angles.
(ii) The linear version of the Kovchegov equation is the BFKL
equation, and when scattering is far from unitarity limits the
Kovchegov equation becomes the BFKL equation.  There appears to be no
limit in which the BMS equation reduces to the BFKL equation.  The
linear limit of the BMS equation is just low orders of perturbation
theory.

Is it just accidental that the BMS equation is the same as an equation
which naturally appears in high-energy scattering, or is it rather an
indication that BFKL dynamics is also present in jet decays, waiting
to be found?  We believe the latter is the case, and we have found an
observable in jet decays which, in certain kinematic regions, is
dominated by BFKL dynamics.  The observable is easiest to describe in
the context of $e^+e^-$ annihilation, although we could equally well
define similar observables in deep inelastic scattering or in
hadron-hadron collision events with large $E_T.$ In the center of mass
of the collision let $E_0$ be the energy of the $e^+e^-$ pair.  The
observable is the number of heavy quark-antiquark pairs produced
having pair mass $\mathcal{M}$ on the order of $2M,$ with $M$ the
heavy quark mass, and with $k_0,$ the energy of the pair, in the
regime $k_0/\mathcal{M}$ on the order of one.  We suppose that
$E_0/k_0 >> 1.$

Of course most heavy quarks are not produced in the kinematic region
where we are looking.  The majority of heavy quarks will be collinear
with one of the two primary jets and their production will be
dominated by double logarithmic terms.  Our choice of kinematics is
motivated by the fact that double logarithmic terms cancel out in the
region we have taken.  In subsequent sections, where the calculation
is performed, we choose, for technical reasons, a boosted frame where
the small angle approximation can be used.

In section 2, we calculate the lowest order contribution of a heavy
quark-antiquark emission from a QCD dipole.

In Section 3, we calculate the evolution of the two-jet system,
originally produced in the $e^+e^-$ collision, in terms of QCD
dipoles.  This section makes use of the large $N_c$ approximation,
however, it may well be that similar results can be obtained without
the large $N_c$ approximation.

In section 4 we compare the two-jet evolution equation with the dipole
version of QCD evolution in high-energy scattering and find them to be
formally equivalent, but in terms of different variables.  In jet
evolution it is the angles of gluons which are frozen during
subsequent emissions and hence become the natural variables, while in
high-energy scattering it is transverse coordinates which are frozen
and so are the natural variables.

In section 5 we give the rate of heavy quark-antiquark pair production
in our chosen kinematic region, and we find that $\alpha_P-1,$ the
hard pomeron intercept, determines the jet energy dependence of the
production.  

\vskip 10pt
\noindent{\bf 2.  The observable and the Born approximation}
\vskip 6pt 

The observable we consider is heavy quark-antiquark production in jet
decay.  The calculation will be done in the large $N_c$ (planar)
approximation where one can view jet evolution in terms of a sequence
of softer and softer gluon emissions which in turn can be viewed as
the production of more and more QCD dipoles.  It may be that much of
what we derive can also be obtained beyond the large $N_c$ limit, but
that is beyond the scope of the present paper.  The simplicitiy of the
large $N_c$ limit is that one may view the gluon as being made of a
quark part and an antiquark part.

In this section we focus on the heavy quark-antiquark pair
$(Q\bar{Q})$ emission from a dipole consisting of the quark part of a
gluon of momentum $p_a$ and the antiquark part of a gluon of momentum
$p_b.$ The gluons $p_a$ and $p_b$ have come from earlier parts of the
decay of a jet, or pair of jets, and it is assumed that $\vert
\vec{p}_a\vert, \vert\vec{p}_b\vert \gg \vert\vec{k}\vert = \ka$ with
$k_\mu$ the four-momentum and ${\sqrt{k_\mu k^\mu}}=\mathcal{M}$ the
mass of the $Q\bar{Q}$ pair.  We denote by $p_1$ and $p_2$ the momenta
of the $Q$ and $\bar{Q},$ respectively, with $k_\mu=(p_1+p_2)_\mu.$ It
is important to emphasize that $k$ is the softest momentum in our
problem, softer than all other gluons emitted in the secondary
branchings, so below we are able to use the soft momentum
approximation.

\begin{center}
\begin{figure}
\centerline{\epsfbox[0 0 213 83]{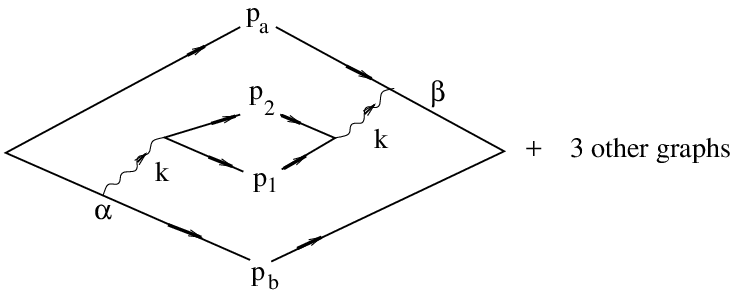}}
\centerline{Fig.~1}
\end{figure}
\end{center}

We now turn to the calculation of the lowest order contribution to
pair production from the dipole $(p_a, p_b).$ We denote this
contribution by ${dN_{a,b}^{(0)}\over d{\mathcal M}^2dy},$ where
$y=\ln \ka/\mathcal{M}.$ One of the lowest order graphs is shown in
Fig.~1 where the left-hand part of the graph denotes the amplitude and
the right-hand part of the graph the complex conjugate part of the
amplitude.  The factors accompanying the coupling of the gluon,
labelled by $k,$ to the ``quark'' $a$ or to the ``antiquark'' $b$ are
given by
\begin{equation}
\label{eq:pab}
g{(2p_a+k)_\alpha\over (p_a+k)^2} \simeq g {p_{a\alpha}\over p_a\cdot k}\,,
\qquad
-g{(2p_b+k)_\alpha\over (p_b+k)^2} \simeq -g{p_{b\alpha}\over p_b\cdot k}\,,
\end{equation}
with color factors to be put in later.  In writing \eqref{eq:pab} we
assume a leading soft approximation where
$\vert\vec{k}\vert/\vert\vec{p}_a\vert$ and $\vert\vec{k}\vert
/\vert\vec{p}_b\vert \ll 1,$ but we do not assume collinear emission.
Indeed, we shall in the end fix ${\mathcal M}/\ka\equiv {\mathcal
  M}/\vert\vec{k}\vert$ so as to guarantee that collinear emission is
unimportant.

One can write
\begin{equation}
\begin{split}
\label{eq:dN0}
{dN_{ab}^{(0)}\over d{\mathcal M}^2dy}= 
& g^2{C_F\over 2} \int{d^4k\over (2\pi)^4{\mathcal M}^4}\, 
\delta\left(k_\mu k_\mu-{\mathcal M}^2\right)
\delta\left(y-\ln  \frac{\ka}{{\mathcal M}}\right)\\
&\left({p_{a\alpha}\over p_a\cdot k} - {p_{b\alpha}\over p_b\cdot
      k}\right) 
\left({p_{a\beta}\over p_a\cdot k} - {p_{b\beta}\over
    p_b\cdot k}\right) 2 {\rm Im}\, \pi^{\alpha\beta}(k)
\end{split}
\end{equation}
where
\begin{equation}
\begin{split}
\label{eq:Imp}
& 2\,{\rm Im}\,\pi_{\alpha\beta}(k) = 
2\,\left(k^2g_{\alpha\beta} - k_\alpha k_\beta\right) {\rm Im}\, \pi(k^2) 
\\&
= \frac{g^2}{(2\pi)^2}\int {d^3p_1d^3p_2\over 2E_1 2E_2}\, 
\delta^4(k\!-\!p_1\!-\!p_2)\, 
{\rm tr}\left\{(\gamma\cdot p_1\!+\!M)
\gamma_\alpha(\gamma\cdot p_2\!-\!M) \gamma_\beta\right\}
\end{split}
\end{equation}
with $M$ the heavy quark mass.  From \eqref{eq:Imp} it is
straightforward to get \cite{xam}
\begin{equation}
\label{eq:Imp'}
{\rm Im}\, \pi({\mathcal M}^2) = - {\alpha_s\over 3} 
{\sqrt{{\mathcal M}^2 -4M^2\over  {\mathcal M}^2} }\ 
{{\mathcal M}^2+ 2M^2\over {\mathcal M}^2}.
\end{equation}
Because of current conservation the $k_\alpha k_\beta$ term in ${\rm
  Im}\,\pi_{\alpha\beta}(k)$ does not contribute to \eqref{eq:dN0}
leaving
\begin{equation}
\begin{split}
\label{eq:dN0'}
{dN_{ab}^{(0)}\over d{\mathcal M}^2dy} =
& -4\pi\alpha_s C_F\int {d^4k\over (2\pi)^4{\mathcal M}^4}
\delta\left(k_\mu k^\mu-{\mathcal M}^2\right)
\delta\left(y-\ln\frac{\ka}{\mathcal M}\right)
\\&
\cdot \frac{p_a\cdot p_b}{(p_a\cdot k)(k\cdot p_b)}
\,2\,{\mathcal M}^2\, {\rm Im}\,\pi({\mathcal M}^2).
\end{split}
\end{equation}
It is now staightforward to get
\begin{equation}
\label{eq:dN0''}
{dN_{ab}^{(0)}\over d{\mathcal M}^2dy} =
{\alpha_s^2 C_F\over 6\pi^3}
\sqrt{\frac{{\mathcal M}^2-4M^2}{{\mathcal M}^2}} 
{{\mathcal M}^2+2M^2\over {\mathcal M}^4} \int 
{\theta_{ab}^2\,d\Omega_k\over 
\left(\theta_{ak}^2+ \frac{{\mathcal M}^2}{\ka^2}\right)
\left(\theta_{kb}^2 +\frac{{\mathcal M}^2}{\ka^2}\right)}
\end{equation}
where $\theta_{ak}$ is the angle between the vectors $\vec{p}_a$ and
$\vec{k},$ with similar definitions for $\theta_{kb}$ and
$\theta_{ab},$ and where we take ${\mathcal M}/\ka \ll 1,$ and we
suppose $\theta_{ab}\ll 1$ so that the small angle approximation 
$\cos\theta_{ab} \simeq 1 - \half\theta_{ab}^2$ along with similar
approximations for $\cos \theta_{ak}$ and $\cos \theta_{kb}$ can be
used. Finally, it is straightforward to evaluate
\begin{equation}
\label{eq:Intheta}
\int\frac{\theta_{ab}^2\, d\Omega_k}
{\left(\theta_{ak}^2 \!+\! \frac{{\mathcal M}^2}{\ka^2}\right)
 \left(\theta_{bk}^2 \!+\! \frac{{\mathcal M}^2}{\ka^2}\right)} = 
{2\pi\,\lambda\over{\sqrt{1\!+\!\lambda^2}}} 
\ln\left[{{\sqrt{1\!+\!\lambda^2}}+\!\lambda\over
          {\sqrt{1\!+\!\lambda^2}}-\!\lambda}\right],
\quad \lambda = {\ka\,\theta_{ab}\over 2{\mathcal M}}.
\end{equation}
This gives
\begin{equation}
\label{eq:dN0-fine}
{d N_{ab}^{(0)}\over d{\mathcal M}^2 dy} = 
{\alpha_s^2 C_F\over 3\pi^2} 
{{\sqrt{{\mathcal M}^2\!-\!4M^2}}\over {\mathcal M}^3} 
\left(1\!+\!{2M^2\over {\mathcal M}^2}\right) 
{\lambda\over {\sqrt{1\!+\!\lambda^2}}} 
\ln\left[{{\sqrt{1\!+\!\lambda^2}}+\!\lambda\over 
{\sqrt{1\!+\!\lambda^2}}-\!\lambda}\right].
\end{equation}
We imagine choosing ${\mathcal M}^2\!-\!4M^2$ on the order of $M^2$ so
that the counting of heavy quarks and heavy mesons should be the same
and physically observable.  We also choose $\lambda$ on the order of
one so that there are no collinear singularities in the emission of
the heavy quark pair from the dipoles.  
\newpage
\vskip 10pt
\noindent{\bf 3. The evolution equation}
\vskip 5pt

The evolution (branching) \cite{tto,ini} of QCD jets has been widely
studied and furnishes some of the best tests of perturbative QCD.
Double logarithmic terms, having both soft and collinear
singularities, dominate the behavior of most global observables, and
the resummation of these perturbation series is now well understood.

The evolution we are going to review in this section is single
logarithmic. Soft, but not collinear singularities, will be summed.
Such terms govern the behavior of certain non-global observables such
as $E_{\rm out},$ the total energy emitted into a region $C_{\rm out}$
away from all the hard jets.  The same single logarithmic terms will
dominate heavy quark-antiquark production in certain regions of phase
space as we shall see below.  However, for the moment we are going to
describe more formally the evolution that will be dominant for our
heavy quark-antiquark production.  We shall see that the evolution is
determined by the formalism developed in by Banfi, Marchesini and Smye
(BMS) \cite{Ban} in their discussion of the distribution function for
$E_{\rm out}$ mentioned above.  We shall also see, perhaps
surprisingly, that the evolution is formally exactly equivalent to
BFKL evolution as described in the dipole formalism.

We begin by considering the process $e^++e^-\rightarrow
\gamma^\ast(q)\rightarrow$ quark $ (p_a) +$ antiquark $(p_b).$ We
chose an unusual frame where $\vec{q}$ is large and along the z-axis.
Without loss of generality we may suppose that $\vert\vec{p}_a\vert =
\vert\vec{p}_b\vert = E$ for our zero mass quarks and that

\begin{equation*}
\theta_{ab}\simeq {{\sqrt{q_\mu  q^\mu}}\over E} = {Q\over E}
\end{equation*}
is small.  $Q,$ the largest physical scale, is assumed to be much
greater than the heavy quark mass $M$ introduced in the previous
section.

Of course, the quark-antiquark pair $(p_a, p_b)$ will further evolve
by gluon emission.  The gluon emission probabilities are known
analytically \cite{tto} in the large $N_c$ limit and in the leading
soft approximation for longitudinal momenta.  No assumption of a
collinear approximation is necessary.  In the large $N_c$
approximation the states having $n$ gluons in addition to the original
quark-antiquark pair may be viewed as a state of $n+1$ dipoles.  For
example, the state having a quark-antiquark pair and a single gluon
consists of two dipoles.  The first dipole is the original quark and
the antiquark part of the gluon while the second dipole consists of
the quark part of the gluon and the original antiquark.  Since our
object is to calculate heavy quark-antiquark production, and since the
heavy quark pair can only be emitted from a single dipole, the same
dipole in the amplitude and in the complex conjugate amplitude, what
we need to extract from the QCD evolution is only the inclusive dipole
distribution caused by the gluon emissions.

Therefore, we need to compute $n(\theta_{ab},\theta,Y)$ the number
density of dipoles of opening angle $\theta,$ starting from the
original quark-antiquark pair which has opening angle $\theta_{ab},$
and in a rapidity interval $Y=\ln E/\ka$. (Here we take the rapidity
of a quark or gluon of energy $\omega$ to be $y=\ln \omega/\Lambda.$)
The dipole having opening angle $\theta$ consists of the quark part of
a gluon, say $g_1,$ and the antiquark part of a gluon, $g_2.$ (The
original quark or antiquark may also make up either the quark or
antiquark part of the dipole $\theta,$ but this is unlikely to be the
case at large $Y.$ Our formalism allows this possibility.)  Let
$\omega$ be the smallest of the energies of the gluons $g_1$ and
$g_2.$ Then in $n(\theta_{ab}, \theta, Y)$ we require that $\ln
E/\omega \ll Y.$ The result for the distribution of heavy quark pairs
is
\begin{equation}
\label{eq:dN}
{dN_{ab}\over d{\mathcal M}^2dY} = 
\int {d\Omega_{b^\prime}\over 4\pi}\,
n(\theta_{ab}, \theta_{a^\prime  b^\prime},Y)\,
{dN_{a^\prime b^\prime}^{(0)}\over d{\mathcal M}^2dy}\,,
\end{equation}
where $y=\ln \kappa/{\cal M}$ and the details of our normalization for
$n$ will be explained later.

In order to write an equation for $n$ we need only an explicit
expression for the emission of a single gluon, of momentum $k,$ from
the original quark-antiquark pair.  By a simple calculation (see for
example, Eqs. 3.2-3.4 of BMS), one finds for the number of gluons
$N_g$
\begin{equation}
\label{eq:dNg}
dN_g= \bar{\alpha}_s\,\omega\, d\omega {d\Omega_k\over 4\pi}\>
{p_a\cdot p_b\over (p_a\cdot k)(k\cdot p_b)}\,,
\qquad
\bar{\alpha}_s={\alpha_s  N_c\over \pi}\,.
\end{equation}
(In covariant gauge only the two graphs shown in Fig.~2 contribute.)
Going to angular variables
\begin{equation}
\label{eq:dNgs}
dN_g = 2\bar{\alpha}_s {d\Omega_k\over 4\pi}\,dy\> 
{\theta_{ab}^2\over \theta_{ak}^2\,\theta_{kb}^2}
\end{equation}
in the small angle approxiation, and where $y=\ln \omega/\Lambda.$

Eq.~\eqref{eq:dNgs} can be turned into an equation for the inclusive
dipole density simply by observing that the measured dipole has as a
parent dipole either a + antiquark part of $k$ or quark part of $k +
b.$ Thus, as illustrated in Fig.~3,
\begin{equation}
\begin{split}
\label{eq:eveq-jet}
{dn(\theta_{ab},\theta,Y)\over dY}
={\bar{\alpha}_s\over 2\pi}\!
\int {\theta_{ab}^2\,d\Omega_k \over \theta_{ak}^2\theta_{kb}^2} 
\,\left[n(\theta_{ak},\theta,Y)\!+\! 
 n(\theta_{kb},\theta,Y)\!-\! 
 n(\theta_{ab},\theta,Y)\right],
\end{split}
\end{equation}
where the third term on the righthand side of \eqref{eq:eveq-jet} is
the virtual contribution of the gluon $k$. Eq.~\eqref{eq:eveq-jet} is
our basic equation, and we shall come back to it when calculating an
explicit asymptotic result for heavy quark-antiquark production.  
{\begin{center}
\begin{figure}[htb]
\centerline{\epsfbox[0 0 252 115]{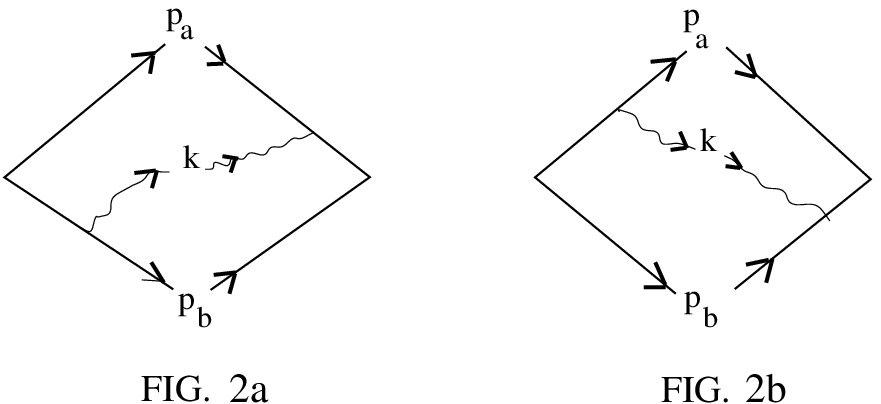}}
\end{figure}
\end{center}
\begin{center}
\begin{figure}[htb]
\centerline{ \epsfbox[0 0 352 66]{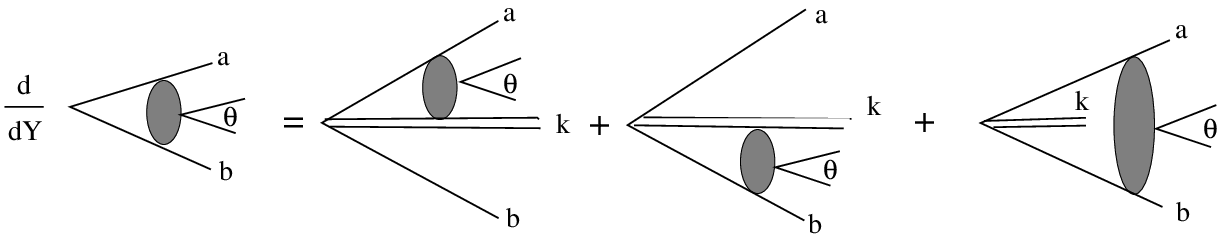}}
\centerline{Fig.~3}
\end{figure}
\end{center}
In Fig.~3 the double line, $k,$ represents the gluon, $k,$ emitted
coherently from $a$ and $b.$ The final term on the righthand side of
Fig.~3 has the $k$-line being emitted and then reabsorbed so that it
does not appear in the final state.

\vskip 10pt
\noindent{\bf 4. Relationship to BFKL dynamics}
\vskip 5pt 

Eq.~\eqref{eq:eveq-jet} bears a remarkable resemblance to the BFKL
equation as given in the dipole formulation \cite{ler,tel,haw}.  The
BFKL equation reads
\begin{equation}
\label{eq:eveq-S}
\begin{split}
{dn(x_{01},x,\!Y)\over dY}
\!=\!{\bar{\alpha}_s\over 2\pi}\!\!
\int\! {x_{01}^2\,d^2x_2\over x_{02}^2 x_{21}^2}\,
\left[n(x_{02},x,\!Y)\!+\!n(x_{21},x,\!Y)\!-\!n(x_{01},x,\!Y)\right],
\end{split}
\end{equation}
for the number density of dipoles of size $x$ to be found in a parent
dipole of size $x_{01}$ over a rapidity range $Y,$ where the parent
dipole is constructed from a quark at $\b{x}_0$ and an antiquark at
$\b{x}_1$ with $x_{01}=\vert\b{x}_0-\b{x}_1\vert$. The measured dipole
has a transverse coordinate separation $x$ between its quark and
antiquark parts.  In \eqref{eq:eveq-S} we anticipate that the
inclusive dipole distribution is independent of the orientations of
the dipoles when $Y$ is large.  (In \eqref{eq:eveq-jet} we have
similarly anticipated that $n$ only depends on the magnitudes
$\theta_{ab}$ and $\theta,$ but not on the orientation between the
angles of the dipoles.)  From \eqref{eq:eveq-S} it is straightforward
to evaluate high-energy dipole-dipole scattering whose rapidity
(energy) dependence is the same as $n$ in \eqref{eq:eveq-S}.

Except for the measure of integration $d\Omega_k$ versus $d^2x_2$
equations \eqref{eq:eveq-jet} and \eqref{eq:eveq-S} are identical.
However, in the
small angle limit,
\begin{equation}
\label{eq:dOm}
d\Omega_k \simeq d\phi_k \theta_k d\theta_k
\end{equation}
can be viewed as an integration measure over a flat plane with
$\theta_k$ and $\phi_k$ the radial and polar coordinates respectively.
Thus \eqref{eq:eveq-jet} and \eqref{eq:eveq-S} are formally identical
equations so long as we allow the $\theta_k$-integration to cover the
region
\begin{equation}
\label{eq:extend}
0 < \theta_k < \infty
\end{equation}
in \eqref{eq:eveq-jet} while still using \eqref{eq:dOm}.  Of course
when $\theta_{ab}$ is small the important region of integration in
\eqref{eq:eveq-jet} is $\theta_k \lesssim \theta_{ab}$ so that
\eqref{eq:extend} is necessary only to make \eqref{eq:eveq-jet} and
\eqref{eq:eveq-S} exactly the same.

Although \eqref{eq:eveq-jet} and \eqref{eq:eveq-S} are formally
indentical the physical variables which appear in these two equations
are very different.  In jet evolution angles seem to be the preferred
variables while in high-energy scattering transverse coordinates seem
to be more natural.  It is not hard to see why this is the case.  In
both cases the basic amplitude is the emission of a softer gluon off
a higher momentum quark or gluon as illustrated in Fig.~4 for a
momentun labelling of these gluons.

\begin{center}
\begin{figure}[htb]
\centerline{\epsfbox[0 0 119 49]{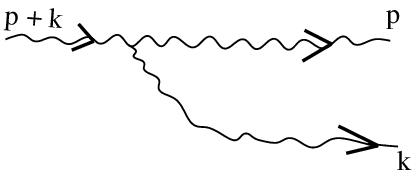}}
\centerline{Fig.~4}
\end{figure}
\end{center}

\noindent
In the high-energy scattering case the graph in Fig.~4 represents a
gluon (p) + gluon (k) part of a light-cone wavefuncton.  The soft
emission limit means $k_+/(p+k)_+=z \ll 1.$ The transverse momenta
$\underline{k}$ and $\underline{p}$ are typically of the same size,
and in the double logarithmic region $\vert\underline{p}+
\underline{k}\vert \ll \vert\underline{k}\vert$. The light-cone
quantization energy denominator is
\begin{equation}
\label{eq:D-S}
D^{-1}\!=\!\left[p_- + k_- -(p\!+\!k)_- \right]^{-1}
\!\!=\!\left[{\underline{p}^2\over 2p_+}\!+\!{\underline{k}^2\over 2k_+}\!-\!
      {(\underline{p}\!+\!\underline{k})^2\over 2(p\!+\!k)_+}\right]^{-1}\!\!
\simeq\left[\underline{k}^2\over 2k_+\right]^{-1}\!\!\!,
\end{equation}
and is dominated by the softest gluon, $k.$ Thus, the time over which
the gluon, $k,$ is being emitted is
\begin{equation}
\label{eq:time}
\tau_k \simeq {2k_+\over \underline{k}^2}\,.
\end{equation}
The transverse velocity of the gluon $p$ is
$\underline{v}_p=\underline{p}/p_+$ while that of the gluon $k$ is
$\underline{v}_k=\underline{k}/k_+$. Over the time during which $k$ is
being emitted the changes in the transverse coordinate positions of
the $p$ and $k$-lines are
\begin{equation}
\label{eq:dxp}
\vert\Delta\,\underline{x}_p\vert\simeq\vert \underline{v}_p\vert\,\tau_k
\>\propto\> {k_+\over p_+}\, \vert\underline{k}\vert^{-1}
\end{equation}
and
\begin{equation}
\label{eq:dxk}
\vert\Delta\,\underline{x}_k\vert \simeq\vert \underline{v}_k\vert\,\tau_k
\>\propto\>\vert\underline{k}\vert^{-1}
\end{equation}
respectively.  Eq.~\eqref{eq:dxk} reflects the uncertainity principle
while \eqref{eq:dxp} says that the harder gluon has a very small
change in its transverse coordinate during the emission of softer
gluons.  It is thus a good approximation to ``freeze'' the transverse
coordinate of a high momentum quark or gluon in a light-cone
wavefunction during the time of formation of the softer parts of the
wavefunction.

On the other hand in jet evolution the energy denominator is better written as
\begin{equation}
\label{eq:D-jet}
D={z(1-z)\over 4}\,(p\!+\!k)_+\,
[\,\underline{\theta}_p-\underline{\theta}_k\,]^2
\end{equation}
with
\begin{equation}
\label{eq:z}
z = {k_+\over (p+k)_+}\,,\qquad  1-z={p_+\over (p+k)_+}
\end{equation}
and
\begin{equation}
\label{eq:thep}
\underline{\theta}_p={\underline{p}\over p_z} = 
{{\sqrt{2}}\, \underline{p}\over p_+},\qquad
\underline{\theta}_k={{\sqrt{2}}\,\underline{k}\over k_+}\,.
\end{equation}
In jet evolution $\underline{\theta}_p$ and $\underline{\theta}_k$ are
typically of the same size, with
$\vert\,\underline{\theta}_p-\underline{\theta}_k\vert
\ll\vert\underline{\theta}_p\,\vert$ in the double logarithmic limit
so that typically $\vert\underline{k}\vert \propto z\,
\vert\underline{p}\vert$ for soft gluon emission.  This means that
\begin{equation}
\label{eq:thpk}
\underline{\theta}_{p+k} = 
{{\sqrt{2}}\, (\underline{p}+\underline{k})\over (p+k)_+} \simeq 
{{\sqrt{2}}\, \underline{p}\over p_+} = \underline{\theta}_p
\end{equation}
for soft gluon emission.  Eq.~\eqref{eq:thpk} indicates that the angle
of a harder gluon does not change due to softer gluon emissions and so
angular variables naturally appear in the jet evolution equation
\eqref{eq:eveq-jet}.

One final issue in comparing high-energy evolution with jet evolution
is the question of the running of the QCD coupling.  Equations
\eqref{eq:eveq-jet} and \eqref{eq:eveq-S} have been written for fixed
coupling evolution.  Running coupling effects come in very differently
in \eqref{eq:eveq-jet} and \eqref{eq:eveq-S}.  In \eqref{eq:eveq-jet},
following BMS, we introduce a variable $\Delta$ where
\begin{equation}
\label{eq:De}
\Delta = \int_{E_0}^E{d\omega\over \omega} 
\bar{\alpha}_s (\omega^2\theta_{ab}^2)\,,
\qquad
{\partial\over \partial\Delta}=
\frac{E\,\partial}{\bar{\alpha}_s\,\partial E}=
{1\over \bar{\alpha}_s} { \partial\over \partial Y},
\end{equation}
and \eqref{eq:eveq-jet} becomes (replacing $n(\theta_{ab},\theta,Y)\to
n(\theta_{ab},\theta,\Delta)$)
\begin{equation}
\label{eq:eveq-D}
{\partial n (\theta_{ab},\theta,\Delta)\over \partial\Delta} 
=\int {d\Omega_k\over 2\pi}\, 
{\theta_{ab}^2\over \theta_{ak}^2\theta_{kb}^2}\,
[\,n(\theta_{ak},\theta,\Delta)\!+\!n(\theta_{kb},\theta,\Delta) \!-\! 
n(\theta_{ab}, \theta, \Delta)\,]\,.
\end{equation}
When $\theta$ and $\theta_{ab}$ are of similar magnitude \eqref{eq:De}
and \eqref{eq:eveq-D} should be adequate to represent running coupling
effects.  We suppose, of course, that $E>E_0$ and that
$E_0\theta_{ab}/\Lambda_{\rm QCD}\gg 1.$ (Note that in jet evolution
the diffusion of angles away from $\theta$ and $\theta_{ab}$ is not a
serious problem in determining the argument of the running coupling
since the $\omega$-dependence in \eqref{eq:De} dominates the variation
of $\bar{\alpha}$ during the evolution).  Thus, in jet evolution
running coupling effects are included, when $\theta_{ab}$ and $\theta$
are not too different, simply by the replacement
\begin{equation*}
\bar{\alpha}_s Y \rightarrow \Delta
\end{equation*}
in going from the fixed coupling to the running coupling case.

Running coupling effects in high-energy BFKL applications are
reasonably well understood \cite{ami,Yu,Cia}.  For our purposes it is
sufficient to observe that for $x_{01}$ and $x$ of comparble size one
may take $\alpha_s^{-1}\simeq \alpha_s^{-1}(x_{01}^2) \simeq - b \ln
(x_{01}^2\Lambda_{\rm QCD}^2)$ so long as $\alpha_s^5\,Y^3\ll 1.$ When
$\alpha_s^5\,Y^3\gg 1$ high-energy evolution and jet evolution will
look quite different.  We note that unitarity effects become important in the regime
$\alpha_s^5\,Y^3\ll1$.

\vskip 10pt
\noindent{\bf 5.  The leading soft approximation result}
\vskip 5pt

It now only remains to combine our lowest order result for heavy
quark-antiquark production from a single dipole with the solution of
\eqref{eq:eveq-jet} for large values of $Y$. For large $Y$ the
solution of \eqref{eq:eveq-jet} is given by the standard BFKL formula
\cite{haw}
\begin{equation}
\label{eq:Mellin}
n(\theta_{ab},\theta,Y)={1\over \theta^2} \int_{-\infty}^\infty 
{d\nu\over 2\pi}\, n_\nu\, e^{2\bar{\alpha}_s\chi(\nu)Y}
\left({\theta_{ab}\over \theta}\right)^{1+2i\nu}
\end{equation}
where
\begin{equation}
\label{eq:chi}
\chi(\nu) = \psi(1) - \half\psi(\half-i\nu) - \half\psi(\half + i\nu)
\end{equation}
and $n_\nu$ determinines the normalization of $n$ at $Y\!=\!0.$ If we take
$n_\nu\!=\!4$ then
\begin{equation}
\label{eq:n0}
n(\theta_{ab}, \theta,0) = {2\over \theta^2} 
\,\delta \left(\ln\frac{\theta_{ab}}{\theta}\right) = {2\over \theta}\, 
\delta (\theta\!-\!\theta_{ab})\,,
\end{equation}
in which case, in a notation and normalization used in \eqref{eq:dN},
\begin{equation}
\label{eq:norm}
\int {d\Omega_{b^\prime}\over 4\pi}\, 
n(\theta_{ab},\theta_{a^\prime b^\prime},Y\!=\!0) = 1\,.
\end{equation}
It is straightforward to find
\begin{equation}
\label{eq:nsol}
n(\theta_{ab}, \theta, Y)\quad {_\simeq\atop{Y \,{\rm large}}}\quad
{\theta_{ab}\over \theta^3} \,
{e^{(\alpha_P\!-\!1)Y}\over{\sqrt{{7\over 2}\pi\bar{\alpha}_s\zeta(3)Y}}}\> 
e^{-\ln^2(\theta_{ab}^2/\theta^2)\over 14\,\bar{\alpha}_s\zeta(3)\,Y}
\end{equation}
with, as usual,
\begin{equation}
\label{eq:aP}
\alpha_P-1 = \bar{\alpha}_s\, 4\,\ln 2\,.
\end{equation}
In case of running coupling evolution, Eq.~\eqref{eq:eveq-D}, one
simply replaces $\bar{\alpha}_s Y$ by $\Delta$ in \eqref{eq:nsol}.
However, in the running coupling case it is difficult to get large
values of $\Delta$ in realistic circumstances.

Now using \eqref{eq:dN0-fine} and \eqref{eq:nsol} in \eqref{eq:dN}
one arrives at the number of heavy quark pairs in the original dipole
to be
\begin{equation}
\label{eq:dNsol}
{dN_{ab}\over d{\mathcal M}^2dY} = 
{\alpha_s^2\over 24}\left({\ka\theta_{ab}\over {\mathcal M}}\right) 
{e^{(\alpha_P-1)Y}\over {\sqrt{\frac{7}{2}\alpha_s\,N_c\,\zeta(3)\,Y}}} 
{{\sqrt{{\mathcal M}^2-4M^2}\over {\mathcal M}^3}} 
\left(1 + {2M^2\over {\mathcal M}^2}\right),
\end{equation} 
where $Y=\ln E/\ka$ and we remind the reader that $\mathcal{M}$ is the
mass of the heavy quark-antiquark pair and $\ka=|\vec{k}|$ the pair
momentum and we suppose that ${\ka\theta_{ab}/\mathcal{M}}$ is not
too different from 1 in order to stay within the (angular) diffusion
radius for BFKL evolution.
\vskip 10pt
\noindent{\bf Acknowledgements}
\vskip 6pt
\indent This work began when G.M. was a visitor at the LPTHE (Paris VI, Jussieu) and A.M. was a visitor at the LPT (Paris XI, Orsay).  We wish to thank Dr. Yuri Dokshitzer (LPTHE) and Dr. Dominique Schiff (LPT) for their hospitality and support during our visits.  We also wish to thank Drs. Yuri Dokshitzer and Gavin Salam for useful discussions concerning this work.

\end{document}